\documentclass[pra,reprint,a4paper,twocolumn,nofootinbib,showpacs]{revtex4-1}
\usepackage{amsmath,amssymb,amsfonts}
\usepackage[dvips]{graphicx}
\usepackage{graphicx}
\usepackage{txfonts}
\usepackage[T1]{fontenc}
\usepackage{textcomp}
\usepackage{graphicx}
\usepackage{epsfig}
\usepackage{upgreek}
\usepackage{braket}
\usepackage{bbold}
\usepackage{psfrag}
\usepackage[unicode]{hyperref}
\hypersetup{
    unicode=true,                                           
    a4paper=true,
    plainpages=false,
    pdffitwindow=true,                                      
    colorlinks=true,                                        
    linkcolor=blue,                                        
    citecolor=blue,                                         
    filecolor=blue,                                        
    urlcolor=blue                                          
}
\newcommand{\eqnrefp}[1]{{[Eq.~(\ref{#1})]}}

\newcommand{\eqnreft}[1]{{Eq.~(\ref{#1})}}

\newcommand{\figreft}[2]{Fig.~\ref{#1}#2}

\newcommand{\figreftfull}[2]{Figure~\ref{#1}#2}
\newcommand{\figrefp}[2]{[Fig.~\ref{#1}#2]}

\newcommand{\secreft}[1]{{Section~\ref{#1}}}
\newcommand{\secrefp}[1]{{(Section~\ref{#1})}}

\newcommand{\eye}{i}
\newcommand{\pa}{\partial}
\newcommand{\xp}{\,\mathrm{e}^}
\renewcommand{\vec}[1]{\boldsymbol{#1}}

\newcommand{\sech}{\mathrm{sech}}
\renewcommand{\sin}{\mathrm{sin}}
\renewcommand{\cos}{\mathrm{cos}}

\renewcommand{\O}{\mathcal{O}}

\begin{document}

\title{Bright matter-wave soliton collisions at narrow barriers}
\author{J. L. Helm}
\author{T. P. Billam}
\author{S. A. Gardiner}
\affiliation{Department of Physics, Durham University, Durham DH1 3LE, United Kingdom}
\date{\today}


\begin{abstract}
We study fast-moving bright solitons in the focusing nonlinear Schr\"{o}dinger equation perturbed by a narrow Gaussian potential barrier. In particular, we present a general and comprehensive analysis of the case where two fast-moving bright solitons collide at the location of the barrier.  In the limiting case of a $\delta$-function barrier, we use an analytic method to show that the relative norms of the outgoing waves depends sinusoidally on the relative phase of the incoming waves, and to determine whether one, or both, of the outgoing waves are bright solitons.  We show using numerical simulations that this analytic result is valid in the high velocity limit: outside this limit nonlinear effects introduce a skew to the phase-dependence, which we quantify. Finally, we numerically explore the effects of introducing a finite-width Gaussian barrier. Our results are particularly relevant, as they can be used to describe a range of interferometry experiments using bright solitary matter-waves. 
\end{abstract}

\maketitle

\section{Introduction\label{intro}}

Bright solitary matter-waves are solitonlike dynamical excitations observed in atomic Bose-Einstein condensates (BECs) with attractive inter-atomic interactions \cite{khaykovich_etal_science_2002, strecker_etal_nature_2002, cornish_etal_prl_2006}.  They are solitonlike in the sense that they propagate without dispersing \cite{morgan_etal_pra_1997}, emerge largely unscathed from collisions with other bright solitary matter-waves and with external potentials \cite{parker_etal_physicad_2008, billam_etal_pra_2011}, and have center-of-mass trajectories which are well-described by effective particle models \cite{martin_etal_prl_2007, martin_etal_pra_2008, poletti_etal_prl_2008}. They derive these solitonlike properties from their analogousness to the bright soliton solutions of the focusing nonlinear Schr\"{o}dinger equation (NLSE), to which the mean-field description of an atomic BEC reduces in a homogeneous, quasi-one-dimensional (quasi-1D) limit. These bright soliton solutions of the 1D focusing NLSE have been extensively explored in nonlinear optics, both in the context of solitons in optical fibers~\cite{zakharov_shabat_1972_russian, satsuma_yajima_1974, gordon_ol_1983, haus_wong_rmp_1996, Helczynski_ps_2000} and as stable structures existing in arrays of coupled waveguides~\cite{Christodoulides_optlett_1988,cohen_prl_2002} which are described by a discretized NLSE. Although the quasi-1D limit is experimentally challenging for attractive condensates \cite{billam_etal_variational_2011}, bright solitary matter-wave dynamics remain highly solitonlike outside this limit \cite{cornish_etal_prl_2006, billam_etal_pra_2011}. Consequently, bright solitary matter-waves present an intriguing candidate system for future interferometric devices \cite{strecker_etal_nature_2002, cornish_etal_physicad_2009, weiss_castin_prl_2009, streltsov_etal_pra_2009, billam_etal_pra_2011, al_khawaja_stoof_njp_2011, martin_ruostekoski_arxiv_2011}.

A key component of a bright solitary matter-wave interferometer is a mechanism to coherently split and recombine bright solitary matter-waves: the collision of a bright solitary wave with a narrow potential barrier is one way to create such a beamsplitter.  Within a quasi-1D, mean-field description of an atomic BEC, collisions of single solitary matter-waves with potential barriers and wells have been extensively studied \cite{kivshar_malomed_rmp_1989, ernst_brand_pra_2010,lee_brand_2006, cao_malomed_pla_1995, holmer_etal_cmp_2007, holmer_etal_jns_2007}, and sufficiently fast collisions with potential barriers have been shown to lead to the desired beamsplitting effect \cite{holmer_etal_cmp_2007, holmer_etal_jns_2007}. When, in nonlinear optics, the soliton exists in an inhomogeneous array of discrete waveguides, the soliton can be reflected, split or captured  at the position of the inhomogeneity~\cite{Krolikowski_joptsocam_1996, Fratalocchi_pre_2006, Konotop_PRE_1996}. This is equivalent, in the continuum limit of an infinite number of waveguides, to splitting a soliton in the GPE at a $\delta$-function ~\cite{Krolikowski_joptsocam_1996} --- a phenomenon which has been called the ``optical axe''~\cite{Helczynski_ps_2000}. Such splitting has been considered in the context of soliton molecule formation \cite{al_khawaja_stoof_njp_2011}, within a mean-field description, and also in the context of many-body quantum mechanical descriptions: in the latter it has been demonstrated that macroscopic quantum superpositions of solitary waves could be created, offering intriguing possibilities for future atom interferometry experiments \cite{weiss_castin_prl_2009, streltsov_etal_pra_2009}. Recently Martin and Ruostekoski, in Ref.~\cite{martin_ruostekoski_arxiv_2011}, considered an interferometer using a narrow potential barrier as a beamsplitter for harmonically trapped solitary waves, based on the particular configuration of a recent experiment ~\cite{nlqugas_confproc_2010}. In particular this work demonstrated that such a potential barrier can also be used to recombine solitary waves, by arranging for them to collide at the location of the barrier. In such collisions, the relative norms of the two outgoing solitary waves was shown to be governed by the phase difference $\Delta$ between the incoming ones.  In the mean-field description the relative norms of the outgoing waves exhibit enhanced sensitivity to small variations in the phase $\Delta$; however, a simulation of the same system including quantum noise, via the truncated Wigner method \cite{blakie_etal_ap_2008}, showed increased number fluctuations that ultimately negated this enhancement \cite{martin_ruostekoski_arxiv_2011}.

\begin{figure}[t]
  \includegraphics{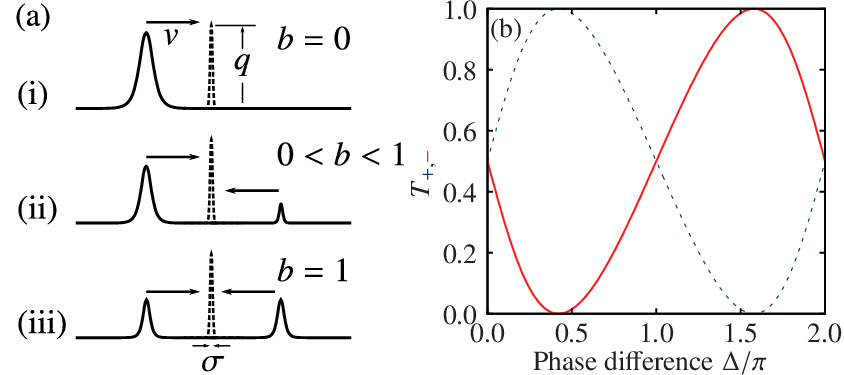}
  \caption{(Color online) (a) Schematic of the collisions we consider: two bright solitons [one in the case $b=0$ (a)(i)] (solid lines) collide at a narrow Gaussian potential barrier (dashed line). The norms of the two outgoing waves are nonlinearly dependent on the relative phase $\Delta$ between the solitons, as illustrated in (b) for equal-amplitude solitons [the case $b=1$ (a)(iii)]; solid red (dashed blue) lines indicate the outgoing wave in the negative (positive) $x$ domain. Here the soliton velocity is $v=2$ and the barrier width is characterised by $\sigma=0.28$}
\label{fig:gauss_nu} 
\end{figure}

In this paper we consider the focusing NLSE perturbed by a narrow, Gaussian potential barrier of the form $V(x) = qe^{-x^2/2\sigma^2}/\sqrt{2\pi}\sigma$, and investigate the dynamics of two fast-moving bright solitons which collide at the location of the barrier. We investigate such collisions for the general initial condition \figrefp{fig:gauss_nu}{(a)} 
  \begin{multline}
    \psi(x) = \frac{1}{2+2b}\left\{\sech\left(\frac{x+x_0}{2+2b}\right)\xp{\eye vx} \right. \\
    \left . +b\sech\left(\frac{b[x-x_0]}{2+2b}\right)\xp{-\eye(vx+\Delta)}\right\},
    \label{icc}
  \end{multline}
with $b > 0$. For large $x_0$ this approximates an exact two-soliton solution comprising \textit{two} bright solitons with \textit{unequal} norms, $1/(1+b)$ and $b/(1+b)$, oppositely directed and equal velocities, $\pm v$, and relative phase $\Delta$ \cite{gordon_ol_1983}.

By examining such collisions for general $b$, $\Delta$, $q$, and $\sigma$ we give a detailed explanation of the nonlinear recombination which occurs after the solitons collide at the potential barrier at time $t_1=2x_0/v$, and are recombined into left- and right-travelling waves in a phase-sensitive way. This general and comprehensive treatment of two-soliton collisions at a barrier constitutes the main result of the paper. For the case of solitons of equal size (as reported in Ref.~\cite{martin_ruostekoski_arxiv_2011}) we illustrate this phase dependence in \figreft{fig:gauss_nu}{(b)}. In this paper we present an analytic description of the recombination for the general two-soliton case ($b>0$) in the limit of a $\delta$-function barrier ($\sigma \rightarrow 0$). This description is derived from an exact description of the single-soliton case ($b=0$) in the same limit \cite{holmer_etal_cmp_2007, holmer_etal_jns_2007}. We compare this to numerical simulations, and find the analytic description is exact in the limit of high velocity. In addition to yielding useful predictions for the relative norms of the recombined waves, this analytic method allows us to estimate whether one, or both, of the outgoing waves are bright solitons. We also numerically investigate the case of a Gaussian barrier, $\sigma > 0$.  Particular cases of interest are $b=0$ ($b\rightarrow \infty$) --- corresponding to a \textit{single} soliton ---  and $b=1$ --- corresponding to \textit{equal}-sized solitons; these correspond, respectively, to the splitting and recombination stages of a bright solitary wave interferometer.  While in the context of atomic BECs the NLSE represents a quasi-1D condensate with tight radial trapping and either zero or very weak axial trapping (e.g., a periodic ``ring'' trap \cite{ramanathan_etal_prl_2011}, or a waveguide or weak harmonic trap \cite{billam_etal_variational_2011}), we emphasise that the equation we study here remains general and could also be used to describe similar systems in, e.g., nonlinear optics. However, as a particular example, our analysis directly allows us to understand the operation of a bright solitary wave interferometer in a ring trap, illustrated schematically in \figreft{fig:int_schem}{(a)}. 

The paper is structured as follows: in \secreft{section:system} we derive the NLSE, perturbed by a narrow Gaussian barrier, in the context of an attractively-interacting atomic BEC. The subsequent sections comprise our analysis of the collisions given by initial condition \eqnreft{icc}. We consider first the single-soliton case ($b=0$), for $\delta$-function  \secrefp{subsec:one_soliton_delta} and Gaussian \secrefp{subsec:one_soliton_Gaussian} barriers, and subsequently the two-soliton case ($b>0$), again for $\delta$-function \secrefp{subsec:two_soliton_analytic} and Gaussian \secrefp{subsec:two_soliton_numerical} barriers. In section \secreft{section:conc} we conclude by interpreting our results in the context of current and future atomic BEC experiments.

\section{Physical system\label{section:system}}

In general a weakly interacting atomic BEC, in the limit of zero temperature, can be described by the 3D Gross-Pitaevskii equation (GPE) \cite{pitaevskii_stringari_2003}:
  \begin{equation}
    \eye\hbar\frac{\pa \Psi(\vec{r})}{\pa t}=\left[-\frac{\hbar^2}{2m}\nabla^2 + V_\mathrm{trap}(\mathbf{r}) + V_\mathrm{ext}(\mathbf{r}) + g_\mathrm{3D}\left|\Psi(\vec{r})\right|^2\right]\Psi(\vec{r}).
  \end{equation}
Here $g_\mathrm{3D}= 4\pi\hbar^2a_sN/m$, and $N$, $m$ and $a_s$ are the atom number, mass, and $s$-wave scattering length respectively. For attractive inter-atomic interactions $a_s<0$. The wave function, $\Psi$, is normalised to 1. The potential $V_\mathrm{trap}(\mathbf{r})=m\omega_r^2(y^2+z^2)/2$ represents the trapping potential, which we take to be a cylindrically symmetric waveguide; such a configuration is approximately achieved in an atomic waveguide trap, or in a toroidal ``ring'' trap \cite{ramanathan_etal_prl_2011} which also introduces periodicity in $x$. 

By increasing the radial trapping one can reach a quasi-1D regime, as defined in detail in Ref.~\cite{billam_etal_variational_2011}, where the radial trapping is tight, but not such that the scattering is no longer  3D [$a_s\ll(\hbar/m\omega_r)^{1/2}$]. In this regime we can separate the radial and axial dynamics with the ansatz $\Psi(\mathbf{r})=\Psi_\mathrm{1D}(x) (m\omega_r/\pi\hbar)^{1/2}\exp{(-m\omega_r[y^2+z^2]/2\hbar)}$. After factoring out global phases associated with the radial harmonic ground state energies, this yields the quasi-1D GPE 
   \begin{equation}
     \eye\hbar\frac{\pa \Psi_\mathrm{1D}(x)}{\pa t}=\left[-\frac{\hbar^2}{2m}\frac{\pa^2}{\pa x^2}+ V_\mathrm{ext}(x) +gN\left|\Psi_\mathrm{1D}(x)\right|^2\right]\Psi_\mathrm{1D}(x).
   \end{equation}
The nonlinearity is quantified by $g=2\hbar \omega_r a_s$. We model the external potential as
  \begin{equation}
    V_\mathrm{ext}(x)=\frac{\hbar\Omega^2}{8\Delta}e^{-2x^2/x_r}.\label{potential}
  \end{equation}
This can be generated by an off-resonant Gaussian light sheet propagating in the $z$ direction with $1/e^{2}$ radii $x_r$ and $y_r$ ($y_r\gg x_r$). In this case $\Delta=\omega_L-\omega_0$ is the detuning of the light sheet's frequency $\omega_L$ from the optical transition frequency $\omega_0$, and $\Omega$ is the Rabi frequency at the centre of the light sheet~\cite{gardiner_etal_pra_2000}.

Working in ``soliton units'' ---  position units of $\hbar^2/mgN$, time units of $\hbar^3/mg^2N^2$, and energy units of $mg^2N^2/\hbar^2$ \cite{billam_etal_variational_2011} --- yields the dimensionless, quasi-1D GPE
  \begin{equation}
    \eye\frac{\pa \psi(x)}{\pa t} =\left[- \frac{1}{2}\frac{\pa^2}{\pa x^2} + \frac{q}{\sigma\sqrt{2\pi}}e^{-x^2/2\sigma^2} - \left|\psi(x)\right|^2\right]\psi(x),
  \end{equation}
where the dimensionless wave function is $\psi = \hbar\Psi_\mathrm{1D}/\sqrt{mgN}$, the normalised barrier width is $\sigma=(\hbar^2/2mgN)x_r$ and the barrier strength is given by 
  \begin{equation}
    q=\frac{x_r\Omega^2\sqrt{2\pi}}{32\omega_ra_sN\Delta}.
  \end{equation}

\section{One-soliton splitting on a narrow barrier ($b=0$)\label{section:one_soliton_splitting}}

\subsection{$\delta$-function barrier ($\sigma \rightarrow 0$)\label{subsec:one_soliton_delta}}

In this section we examine the splitting of a single bright soliton ($b=0$) on a $\delta$-function barrier. The assumption of a $\delta$-function barrier facilitates an analytic treatment and is valid for narrow barriers with $\sigma \rightarrow 0$. A detailed analytic treatment single-bright-soliton splitting on such a barrier is given by Holmer, Marzuola and Zworski in Ref.~\cite{holmer_etal_cmp_2007}. Here we briefly restate two key results of Ref.~\cite{holmer_etal_cmp_2007} within our own notation. 

Firstly, the transmission coefficient for a fast-moving bright soliton splitting on a $\delta$-function barrier is approximately equal to the transmission coefficient for plane waves incident on an identical $\delta$-function barrier in linear quantum mechanics, $T_q(v)$, given by
  \begin{equation}
    T_q(v)=|t_q(v)|^2 =\frac{v^2}{v^2+q^2}=\frac{1}{1+\alpha^2}.
    \label{deltrans}
  \end{equation}
Here, $t_q(v)$ is the transmission amplitude associated with a $\delta$-function barrier in linear quantum mechanics, and the soliton velocity  $v$ plays a role analogous to the wavenumber of the incident wave. The transmission and reflection amplitudes $t_q(v)$ and $r_q(v)$, are defined as
  \begin{equation}
      t_q(v)=\frac{\eye v}{\eye v -q} \qquad\mbox{and}\qquad 
      r_q(v)=\frac{q}{\eye v -q}.
      \label{lin_trans}
  \end{equation}
The quantity $\alpha$ characterises the transmission in the linear case, and hence the transmission of bright solitons in the high velocity limit. The exact relation between $T_q(v)$ and the actual transmission coefficient for the incident bright soliton, 
  \begin{equation}
    T_q^s(v) = \lim_{t\rightarrow \infty} \int_0^\infty|\psi(x,t)|^2 dx,
  \end{equation}
is determined in Ref.~\cite{holmer_etal_cmp_2007} to be
  \begin{align}
    T_q^s(v) &= \frac{v^2}{v^2+q^2}+\O(v^{1-3\eta/2})\notag\\
           &= T_q(v)+\O(v^{1-3\eta/2}),\qquad\mbox{as }v\to\infty,
    \label{thm1}
  \end{align}
provided that the initial offset is $x_0\leq-v^{1-\eta}$ and $\alpha=q/v$ is fixed. Here, $\eta$ is a parameter linked to the duration for which the soliton interacts with the barrier, and must satisfy $2/3\textless\eta\textless 1$. The brevity of this duration for a fast-moving bright soliton, which allows one to treat the splitting as a \textit{linear} process, is fundamental to the proof of the above result \cite{holmer_etal_cmp_2007}. The error term in \eqnreft{thm1} is minimized for brief collisions ($\eta \rightarrow 1$), in which case it decays with velocity as $v^{-1/2}$.

Secondly, it is also determined in Ref.~\cite{holmer_etal_cmp_2007} that the outgoing waves resulting from splitting a bright soliton on a $\delta$-function barrier are composed of either one, or two, bright solitons, and a time-decaying radiation term. This is significant, as previously the transmitted and reflected waves were considered to be only `soliton-like'~\cite{cao_malomed_pla_1995,holmer_etal_jns_2007}. The resulting bright solitons are described, for high velocity, by
  \begin{equation}
    \psi(x,t)=\psi_T(x,t)+\psi_R(x,t)+\O\left(\left[t-|x_0|/v\right]^{-1/2}\right)+\O(v^{1-3\eta/2})\label{thm3}
  \end{equation}
where
  \begin{align*}
    \psi_T(x,t)= \xp{i\varphi_T}\xp{i( xv+[A_T-v^2]t/2)}A_T\sech(A_T[x-x_0-tv]),\\
    \psi_R(x,t)= \xp{i\varphi_R}\xp{i(-xv+[A_R-v^2]t/2)}A_R\sech(A_R[x+x_0+tv]).
  \end{align*}
The amplitudes of the transmitted and reflected solitons are given by
  \begin{equation}
    A_T= \mbox{max}(0,2|t_q(v)|-1) \;\mbox{and}\; A_R=\mbox{max}(0,2|r_q(v)|-1);\label{solamp}
  \end{equation}
in the case that $A_T$ ($A_R$) is equal to zero, the transmitted (reflected) outgoing wave does not contain a soliton, but only radiation. More generally, the inequalities $A_T < T_q^s(v)$ and $A_R < 1-T_q^s(v)$ hold. The phases imparted by the splitting process are defined by
  \begin{align}
  \begin{split}
    \varphi_T&=\arg(t_q(v))+\varphi_0(|t_q(v)|)+[1-A_T^2]|x_0|/2v,\\
    \varphi_R&=\arg(r_q(v))+\varphi_0(|r_q(v)|)+[1-A_R^2]|x_0|/2v,
    \label{stc}
  \end{split}
  \end{align}
where
  \begin{equation}
    \varphi_0(\omega)=\int_0^\infty\ln\left(1+\frac{\sin^2(\pi\omega)}{\cosh^2(\pi\zeta)}\right)\frac{\zeta}{\zeta^2+(2\omega-1)^2}d\zeta.
  \end{equation}

\subsection{Gaussian barriers ($\sigma > 0$)\label{subsec:one_soliton_Gaussian}}

 \begin{figure}[t]
    \includegraphics{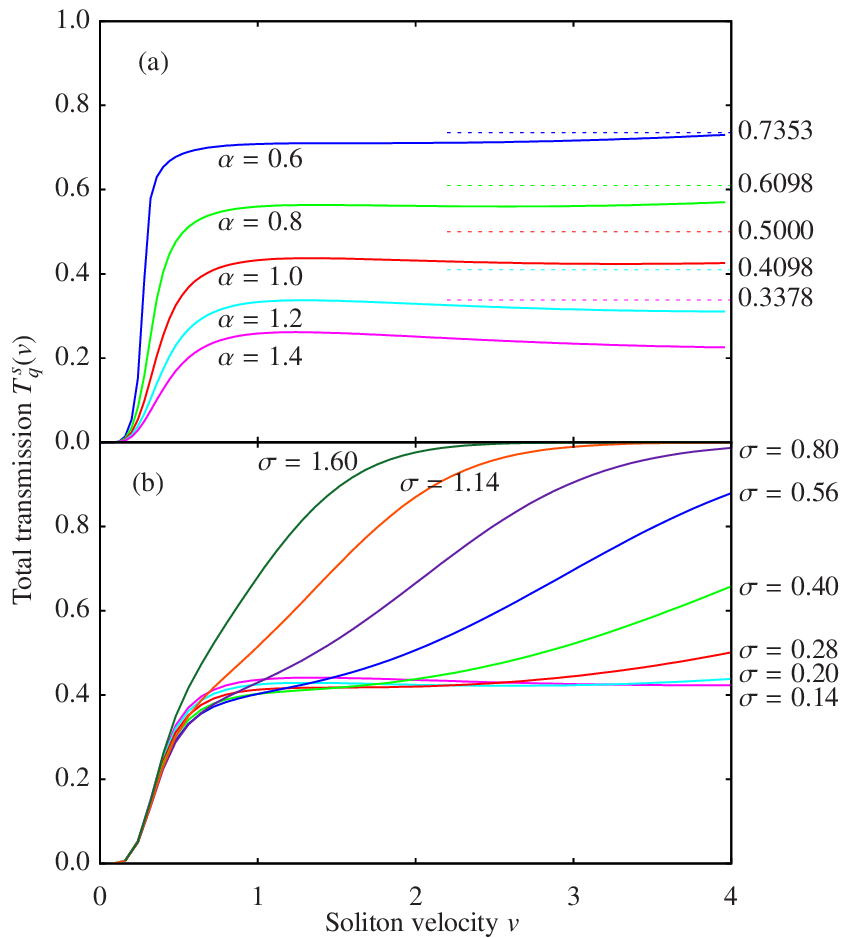}
    \caption{(Color online) (a) Plot of numerically obtained bright soliton transmission, $T_q^s(v)$, as a function of velocity $v$ for a range of fixed $\alpha=q/v$ and a narrow Gaussian barrier with width $\sigma=0.1$. Dashed lines show the transmission through a $\delta$-function in the linear regime for the same range of $\alpha$. (b) Numerically obtained bright soliton transmission for $\alpha=1$ and with a range of barrier widths $\sigma$.\label{fig:gauss_trans}}
 \end{figure}

We now analyse, numerically, the bright soliton splitting process at a Gaussian barrier. Our numerical simulations use a Fourier pseudospectral split-step method with a periodic grid. We ensure that grid size and spacing are chosen such that the bright solitons are well separated and the effects of the periodicity are negligible.

Our initial condition takes the form
  \begin{equation}
    \psi(x)=\frac{1}{2}\xp{\eye vx}\sech([x-x_0]/2),
    \label{gaus1ic}
  \end{equation}
where $x_0<0$. \figreftfull{fig:gauss_trans}{(a)} shows the transmission coefficient $T_q^s(v)$ obtained from numerical simulations of a single bright soliton splitting on a Gaussian barrier with width $\sigma=0.1$, and with $\alpha=q/v=0.6$, $0.8$, $1.0$, $1.2$, and $1.4$.  In our numerics we define $T_q^s(v)$ by the integral of $\psi(x,t_1)$ over the positive $x$ domain,
  \begin{equation}
    T_q^s(v)= \int_0^\infty|\psi(x,t_1)|^2 dx.
  \end{equation}
Here $t_1 = 2|x_0|/v$, such that at this time an unimpeded bright soliton would have reached the point $x=+|x_0|$; at this time the outgoing waves are well-separated. The results are comparable to the $\delta$-function barrier case explored in Ref.~\cite{holmer_etal_cmp_2007} and the previous section.

\figreftfull{fig:gauss_trans}{(a)} shows that as $\alpha$ increases so does the discrepancy between the asymptotic $\delta$-function limit and $T_q^s(v)$. This can be understood by considering how the strength of the barrier compares to the (particle-like) kinetic energy of the soliton $v^2/2$. In the region where the strength of the barrier is greater than the soliton's kinetic energy the wave function decays, reducing transmission. By equating these two values,
  \begin{equation}
    \frac{v^2}{2}=\frac{q}{\sigma\sqrt{2\pi}}\xp{-x_d/2\sigma^2},
  \end{equation}
we determine that the distance over which the wave function decays, $x_d$, is described by
  \begin{equation}
    x_d^2=2\sigma^2\ln\left(\sqrt{\frac{2}{\pi}}\frac{\alpha}{|v|\sigma}\right).
  \end{equation}
It is clear that, for a given $v$ and $\sigma$, as we increase $\alpha$ (by increasing $q$) we increase $x_d$. This is inconsistent with the assumption of a brief barrier-soliton interaction period, which is required in the delta function case of soliton splitting. This insconsistency causes an increase in the attenuation of the wave function, reducing  transmission.

We show the computed dependence of the transmission on the barrier width $\sigma$ in \figreft{fig:gauss_trans}{(b)}. These computations were carried out with $\alpha=1$. For wider barriers or in the higher velocity range, where the peak height of the potential is less than the (particle-like) kinetic energy $v^2/2$ of the incident soliton, the amount of transmission is greatly increased. This illustrates the classical transmission regime where the soliton simply passes through the potential, and, for the Gaussian barriers considered, boils down to an argument that we must have 
  \begin{equation}
    \frac{v^2}{2}\ll\frac{q}{\sigma\sqrt{2\pi}}\Rightarrow\left|v\right|\ll\frac{2\alpha}{\sigma\sqrt{2\pi}}
  \label{qreg}
  \end{equation}
to be definitely out of the classical transmission regime. From \eqnreft{qreg} it is apparent that for satisfactorily large $v$ we will always enter the classical transmission regime for any given finite Gaussian barrier. This regime cannot be retrieved in the $\delta$-function case.

The comparison to the $\delta$-function case is valid in the quantum transmission regime, where the velocity is low enough (for a given $q$,~$\sigma$) that the soliton cannot classically pass through the barrier and must tunnel through instead. For example, this is true when $0.5 \lesssim v \lesssim 2$ and $\sigma\le0.28$~\figrefp{fig:gauss_trans}. Within the quantum transmission regime \eqnrefp{qreg} the $\delta$-function limit of $0.5$ is reached (from below) by reducing $\sigma$. This allows for larger values of $v$, as is consistent with Holmer and Marzuola's work in~\citep{holmer_etal_cmp_2007} where results are general for any $v\gtrsim1$~(and so is in the high velocity regime). 

 \figreftfull{fig:gauss_trans}{(b)} shows that the transmission approaches the analytic prediction for a $\delta$-function barrier as the barrier width $\sigma$ tends to zero. This confirms that the analytic expressions given in Ref.~\cite{holmer_etal_cmp_2007} and the previous section for the $\delta$-function barrier can be quantitatively useful for realistic Gaussian barrier widths. For example, \figreft{fig:gauss_trans}{(b)} indicates the analytic prediction is reasonably quantitatively accurate for $\sigma\le0.28$ in soliton units. For a condensate of $^{85}$Rb and using typical experimental parameters of $N\sim6\times10^3$ atoms, $a_s\sim 5a_0$ (the Bohr radius) and $\omega_r\sim17$Hz this translates to a splitting beam with a full width at half maximum of $\sim9$~$\mu$m. These parameters are consistent with the experimental setup in ~\cite{cornish_etal_prl_2006}. For a similarly sized condensate of $^7Li$ atoms tuned to a similar scattering length this width becomes $\sim2$~$\mu$m. This parameter regime is consistent with \cite{khaykovich_etal_science_2002} appart from the radial trapping frequency, which we reduced from $2\pi \times 710$Hz to $2\pi \times 200$Hz.

\section{Two-soliton collisions at narrow barriers ($b>0$)\label{section:two_soliton_collision}}

\subsection{Analytic treatment for $\delta$-function barrier ($\sigma \rightarrow 0$)\label{subsec:two_soliton_analytic}}

We now give an approximate analytical description of the dynamics of two fast-moving bright solitons colliding at a $\delta$-function barrier, which we subsequently compare to numerical simulations in order to give a fuller picture of the real dynamics that we might expect to see in an experiment. This analysis stops short of the full analytic rigor used in \cite{holmer_etal_cmp_2007} but is consistent within its assumptions of linearity. As previously stated, during the time over which \textit{one} bright soliton interacts with the potential we can describe the system as linear \cite{holmer_etal_cmp_2007}. Here we extend this argument to a scenario in which \textit{two} bright solitons collide at a $\delta$-function potential, as described by the equation
  \begin{equation}
    \eye\frac{\pa \psi(x,t)}{\pa t} =\left[- \frac{1}{2}\frac{\pa^2}{\pa x^2} +q\delta(x) - \left|\psi(x,t)\right|^2\right]\psi(x,t),
  \end{equation}
and initial condition 
  \begin{align}
  \begin{split}
    \psi(x,0)&=\psi_+(x)+\psi_-(x),\\
    \psi_-(x)&=\frac{1}{2+2b}\sech\left(\frac{x+x_0}{2+2b}\right)\xp{\eye vx},\\
    \psi_+(x)&=\frac{b}{2+2b}\sech\left(\frac{b(x-x_0)}{2+2b}\right)\xp{-\eye[vx+\Delta]},
  \end{split}
  \end{align}
\eqnrefp{icc}. We achieve this by making use of the second result of Ref.~\cite{holmer_etal_cmp_2007}, which we apply to the positive and negative domain bright solitons, $\psi_+$ and $\psi_-$, separately, before taking a linear combination of the results. This means that at some time $|x_0|/v\textless t\textless v^{-\eta}+|x_0|/v$ after the barrier collision the solution can be written as a sum of four $\sech$ profiles, two in each of the positive and negative domains;
  \begin{equation}
    \psi(x,t)=\psi_{+T}(x,t)+\psi_{-R}(x,t)+\psi_{-T}(x,t)+\psi_{+R}(x,t).\label{2soloutgoing}
  \end{equation}
Here $\psi_{+T}$ denotes the bright soliton transmitted to the negative domain which originated in the positive domain, $\psi_{-R}$ denotes the bright soliton originating from and reflected back into the negative domain, and so on. In this scheme
  \begin{align}
    \psi_{+T}(x,t)&=\xp{\eye\left(\phi_{+T}+\varphi_{+T} + \Delta\right)}A_{+T}\sech\left(A_{+T}\left[x-x_0+tv\right]\right),\notag\\
    \psi_{+R}(x,t)&=\xp{\eye\left(\phi_{+R}+\varphi_{+R} + \Delta\right)}A_{+R}\sech\left(A_{+R}\left[x+x_0-tv\right]\right),\notag\\
    \psi_{-T}(x,t)&=\xp{\eye\left(\phi_{-T}+\varphi_{-T}         \right)}A_{-T}\sech\left(A_{-T}\left[x+x_0-tv\right]\right),\notag\\
    \psi_{-R}(x,t)&=\xp{\eye\left(\phi_{-R}+\varphi_{-R}         \right)}A_{-R}\sech\left(A_{-R}\left[x-x_0+tv\right]\right).\notag\\
  \end{align}
Two phase factors appear above; the $\phi_{\pm R/T}$ are those associated with the standard soliton solution and are given by
  \begin{align}
  \begin{split}
   \phi_{\pm T}&=\mp vx+\left[A_{\pm T}^2-v^2\right]t/2,\\
   \phi_{\pm R}&=\pm vx+\left[A_{\pm R}^2-v^2\right]t/2.\\
  \end{split}
  \end{align}
The $\varphi_{\pm R/T}$ factors are imparted by the collision, and are described by
  \begin{align}
  \varphi_{\pm T}&=\left[1-A_{\pm T}^2\right]|x_0|/(\mp 2v)+\arg\left(t_q(v)\right)+\varphi_0(|t_q(\mp v)|),\notag\\
  \varphi_{\pm R}&=\left[1-A_{\pm R}^2\right]|x_0|/(\mp 2v)+\arg\left(r_q(v)\right)+\varphi_0(|r_q(\mp v)|).\notag\\
  \end{align}

With $b=1$, barrier height $q=v$, and fast-moving solitons ($v$ large) both initial bright solitons are split equally, such that the amplitudes $A_{\pm R/T}$ are all equal and global phases can be dropped. In this case \eqnreft{2soloutgoing} simplifies dramatically, and shortly after the collision can be written as
  \begin{align}
  \begin{split}
    \psi(x,t)&=\psi_{++}(x,t)+\psi_{--}(x,t)),\label{iccp2}\\
    \psi_{++}(x,t)&=P_{+}(\Delta)f_{+}(x,t),\\
    \psi_{--}(x,t)&=P_{-}(\Delta)f_{-}(x,t),
  \end{split}
  \end{align}
where the terms
  \begin{align}
  \begin{split}
    P_{-}(\Delta)&=\frac{1}{2}\left\{\xp{i\arg(r_q(q))}+\xp{i\left[\arg(t_q(q))+\Delta\right]}\right\},\\
    P_{+}(\Delta)&=\frac{1}{2}\left\{\xp{i\arg(t_q(q))}+\xp{i\left[\arg(r_q(q))+\Delta\right]}\right\},
  \end{split}
  \end{align}
contain information about the constructive and destructive interference between the transmitted and reflected waves. It should be noted that this treatment allows us to infer the bright soliton interactions, but does not give us a complete solution; the terms $f_+$ and $f_-$ contain information about the outgoing wave profiles. By taking a linear superposition of the resultant bright solitons we initially obtain a $\sech$ profile which is not a single-soliton solution. However, in subsequent nonlinear evolution this profile returns to a soliton profile to within a known error, as documented in Appendix B of Ref.~\cite{holmer_etal_cmp_2007}.

  \begin{figure}[t]
    \includegraphics{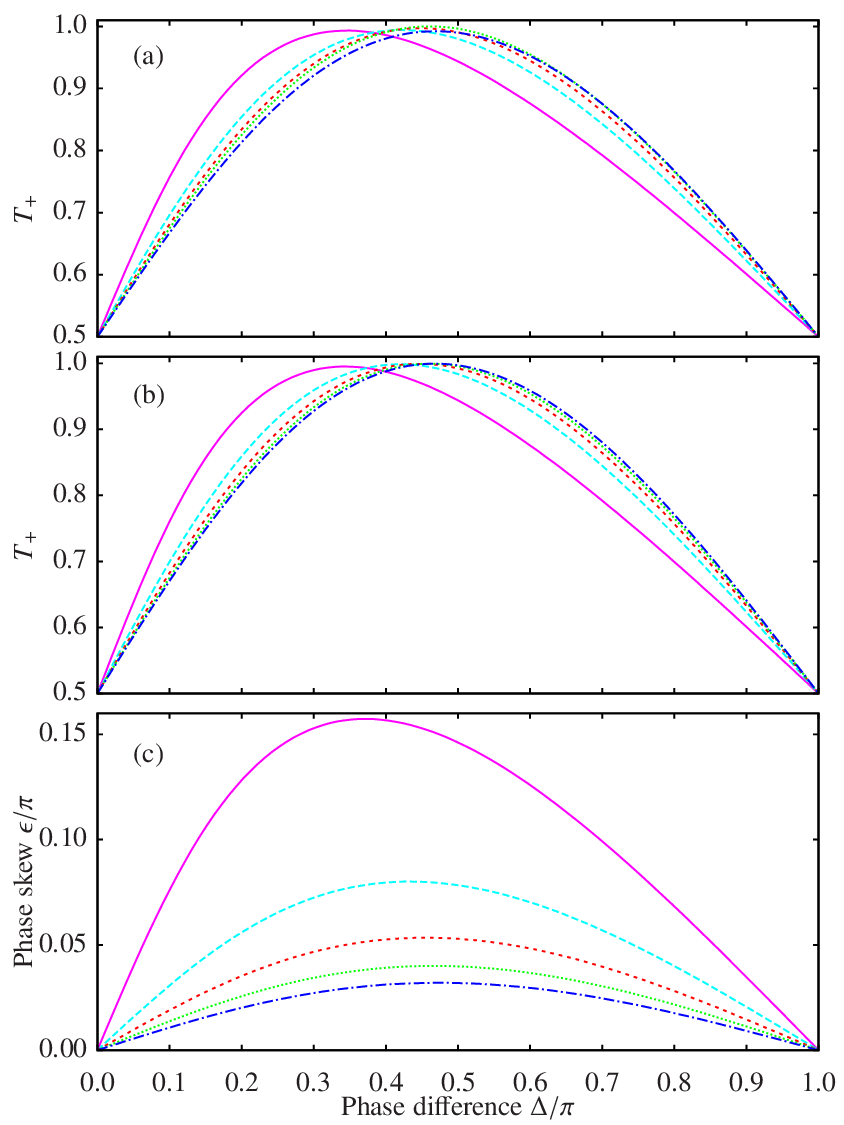}
\caption{(Color online) Phase skew of numerical results with respect to analytic prediction for equal-size ($b=1$) bright soliton collisons at a narrow barrier. (a) Numerically obtained data showing the dependence of the norm of the outgoing wave in the positive domain, $T_+$, after Gaussian barrier collision, on phase difference $\Delta$. Shown here are interference curves for solitons moving with velocity $v=1$ (fuchsia, solid), $2$ (light blue, long dashed), $3$ (red, short dashed), $4$ (green, dotted) and $5$ (dark blue, dot-dashed). The width of the barrier was $\sigma=0.14$. (b) (see (a)) Collisions at a $\delta$-function barrier. Notice the qualitatively identical form of the curves, illustrating that both $\delta$-function and Gaussian barriers exhibit the same skew, and so both undergo the same non-linear effects. (c) Numerically obtained data showing the phase perturbation $\epsilon$~\eqnrefp{epsilondef}~due to non-linear effects in a soliton collision at a $\delta$-function barrier. Shown here, in order of descending amplitude, are the skewness parameters ($\epsilon$) of the interference curves for solitons moving with velocity $v=1$, $2$, $3$, $4$, and $5$ colliding at a $\delta$-function barrier. (a)-(c) are the upper left quarters of the full data set; the plots are both symmetric about the lower and right hand axes. All results shown are calculated for $\alpha=q/v=1$.\label{fig:delta_col_nu}}
  \end{figure}

At a suitably large time after the collision, when the solitons have again separated to the extent that they are again effectively independent, inspection of $|\psi|^2$ shows that the bright solitons are modulated by the factors 
  \begin{align}
  \begin{split}
    |P_-(\Delta)|^2&=\frac{1}{2}\left[1-\sin\left(\Delta\right)\right],\\
    |P_+(\Delta)|^2&=\frac{1}{2}\left[1+\sin\left(\Delta\right)\right].
    \label{pco}
  \end{split}
  \end{align}
These factors determine the norm of the outgoing waves in the positive and negative domains, defined by
  \begin{equation}
    T_{\pm} = \pm \lim_{t\rightarrow \infty} \int_0^{\pm \infty} |\psi(x,t)|^2 dx = |P_{\pm}(\Delta)|^2.
  \end{equation}
Within the analytic approach presented here $T_{\pm}$ are functions of $\Delta$ alone. It should be noted that the symmetry of the initial condition and linear interaction means that the phase interactions apply to both the transmitted and reflected bright solitons and the radiation terms. As a result the quantity $T_\pm$ scribes the total density in the positive and negative domains, not just the respective bright solitons. For suitably high incident velocities this radiation becomes negligible, in accordance with \eqnreft{thm3}.

\subsection{Numerical treatment for $\delta$-function and Gaussian barriers (general $\sigma$)\label{subsec:two_soliton_numerical}}

In \figreft{fig:delta_col_nu}{} we present results of numerical simulations of fast ($v\gtrsim1$) bright soliton collisions at both $\delta$-function\footnote{Within our Fourier pseudospectral method a $\delta$-function barrier can be implemented with high accuracy in momentum space using the approach outlined in Ref.~\cite{sacchetti_jcp_2007}.} and Gaussian barriers. The norms of the outgoing waves, defined in our numerics by
  \begin{equation}
    T_\pm= \pm \int_{0}^{\pm\infty}|\psi(x,t_1)|^2dx,
  \end{equation}
agree qualitatively with the predictions of our analytic treatment, but with a noticeable skew in the predicted sinusoid. This skew is also visible in the results for the Gaussian barrier case shown in \figreft{fig:gauss_nu}. We parametrise this skew by $\epsilon$ and describe the norms of the outgoing waves, $T_{\pm}$, as
  \begin{equation}
    T_\pm = \frac{1\pm\sin(\Delta+\epsilon)}{2}\label{epsilondef}.
  \end{equation}
This skewness parameter is less pronounced for increasing velocities, i.e.,
\begin{equation}
  \lim_{v\to\infty} \mbox{max}(\epsilon) =0.
\end{equation}

The presence of the skew in simulations with both Gaussian and $\delta$-function barriers rules out any explanation in terms of the barrier structure. However, it is well known that when solitons collide in the absence of a barrier they induce a small phase and position shift in one another~\cite{drazin_johnson, al_khawaja_stoof_njp_2011,gordon_ol_1983}. We propose that the skew is a result of interactions between the solitons while approaching the barrier; more fundamentally, this is a result of the condition of a brief interaction not being fully satisfied.  For instance, from initial condition \eqnreft{icc} the phase ($\varphi^\prime_l$) and position ($x^\prime_l$) shift on the left hand soliton are given by
  \begin{equation}
    \frac{2x^\prime_l}{1+b} + \eye\varphi^\prime_l=2\ln\left(\frac{v+\eye}{v+\eye\left[(1-b)/(1+b)\right]}\right).
  \end{equation}
In the case of equal amplitudes and velocities total phase difference reduces to $\varphi^\prime=\pm4\arctan(1/v)$ or, in the limit of high velocity, $\varphi^\prime \approx \pm4/v$. In our scenario only part of this phase-shift can occur before the solitons enter the linear regime, and so we expect that our skewness parameter $\epsilon$ will be some fraction of $\varphi^\prime$. What we have observed from our numerics is that $\epsilon$ oscillates with $\Delta$ but the maximum value is $\epsilon_{\mbox{\scriptsize max}}\approx\varphi^\prime/8$. This is consistent with the behaviour we observe in the high velocity limit.

\begin{figure}[t]
    \includegraphics{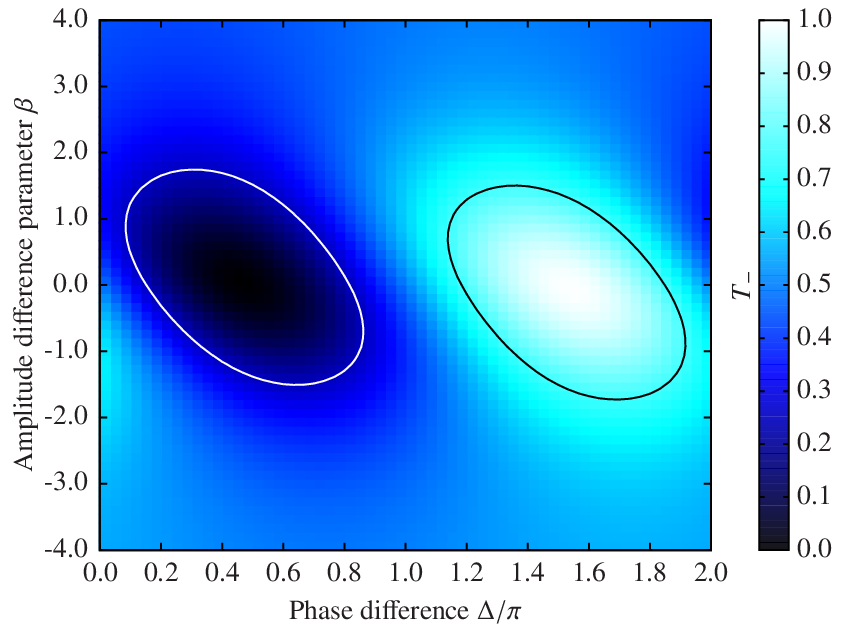}
\caption{(Color online) Numerically computed transmission coefficient $T_+$ illustrating the interference between solitons of different initial amplitudes ($b=\xp{\beta}$) colliding at a $\delta$-function barrier. Even in the case of a large difference in initial amplitude (large $|\beta|$) there is still interference between the solitons. The contour lines show the boundary between having one (interior regions) and two (exterior region) outgoing bright solitons in the analytic treatment [see \eqnreft{solamp}]. All results shown are calculated for $\alpha=q/v=1$ and $v=5$.\label{fig:delta_amp}}
  \end{figure}

It should also be noted that the interference effect is present in collisions between solitons of differing amplitudes. By taking $b =\xp{\beta} $ we see that there is still interference between the transmitted positive and reflected negative bright solitons (and vice versa) \figrefp{fig:delta_amp}{}. Along the line $\beta=0$, where the amplitudes of the incoming bright solitons are equal, we can clearly see a sinusoidal dependence on $\Delta$. For nonzero $\beta$ there is still a notable dependence on the incoming phase difference, but this effect is soon washed out if the difference in initial amplitudes becomes too large. It is true, however, that the solitons do not have to be of similar size to constructively or destructively interfere.

The black (white) contour on \figreft{fig:delta_amp}~shows where the final population in the positive (negative) domain was not great enough, after interference, for the final aggregation to form a soliton. This is determined by treating the total positive (negative) domain population as the transmission (reflection) coefficient in equation \eqref{solamp}. From \eqnreft{solamp} we see that $|t_q|$ and $|r_q|$ must both be $\textgreater0.5$ to get two outgoing solitons.  We determine $|t_q|$ and $|r_q|$ numerically as 

  \begin{equation}
    \sqrt{T_\pm}=\sqrt{\int_0^{\pm\infty}|\psi|^2dx}. 
  \end{equation}

As such, the white contour marks where $T_-=0.25$ ($T_+=0.75$) and the black contour marks where $T_+=0.25$ ($T_-=0.75$).

\section{Applications and conclusions\label{section:conc}}

\begin{figure}[t]
  \includegraphics{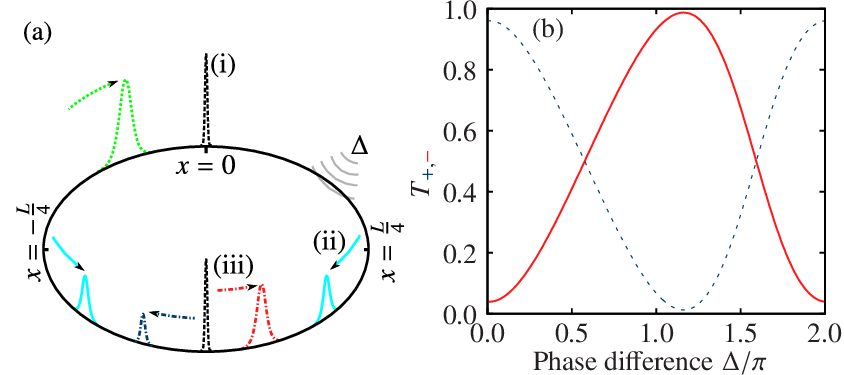}
  \caption{(Color online) (a) Schematic of a ring-trap interferometer; (i) an incoming bright soliton (dotted, green) is split into two equal-amplitude solitons at the first narrow barrier (dashed, black) gaining relative phase $\sim\pi/2$; (ii) these solitons (solid, cyan) propagate around the ring, accumulating an additional relative phase difference $\Delta$; (iii) at the second narrow barrier (dashed, black) these solitons are recombined into outgoing waves [dot-dashed, red (blue) for positive (negative) $x$ domain]. The norms of the two outgoing waves are shown as a function of $\Delta$ in (b), and illustrate the shift by $\sim\pi/2$ with respect to \figreft{fig:gauss_nu}(b)  caused by the initial splitting. Here the soliton velocity is $v=2$ and the barrier width is characterised by $\sigma=0.28$}
\label{fig:int_schem} 
\end{figure}

As stated in \secreft{intro}, an important aspect of our analysis is that is directly allows us to describe the operation of the bright solitary matter-wave Mach-Zender interferometer in a ring trap shown in \figreft{fig:gauss_nu}{(c)}. In the quasi-1D limit we have considered in this paper, such an interferometer can be described by
  \begin{multline}
    \eye\frac{\pa \psi(x)}{\pa t} =\Bigg\{-\frac{1}{2}\frac{\pa^2}{\pa x^2} +\frac{q}{\sigma \sqrt{2\pi}}\Big[ e^{-x^2/2\sigma^2}  \\
    + e^{-(x-L/2)^2/2\sigma^2} \Big]-\left|\psi(x)\right|^2\Bigg\}\psi(x)\label{ring},
  \end{multline}
where $x \in (-L/2,L/2]$ is now a periodic coordinate. In \eqnreft{ring} there are two narrow Gaussian potential barriers, located on opposite sides of the ring trap. Our analysis can be applied to understanding such an interferometer by splitting its operation into the following three stages (illustrated in \figreft{fig:int_schem}):

Firstly, a single initial bright soliton is incident on the barrier at $x=0$~(\figreft{fig:int_schem}{(a)(i)}), at high velocity. Assuming an initial displacement of $x_0=L/4$ is sufficient for the soliton to be well-separated from the barrier, and a barrier height $q=v$ (such that $\alpha = 1$), the analysis of \secreft{section:one_soliton_splitting} applies, with $b=0$. Hence, we obtain two equal-sized outgoing bright solitons, with relative phase $\pi/2$.

Secondly, these bright solitons propagate without dispersion in opposite directions around the ring. We assume that the soliton in the positive $x$-domain picks up an additional phase shift $\Delta$ due to the effects of whatever interaction the interferometer is measuring \figrefp{fig:int_schem}{(a)(ii)}.

Thirdly, these bright solitons collide at the barrier at $x=L/2$. Here, the analysis of \secreft{section:two_soliton_collision} applies, with $b=1$, $\alpha=1$, and $\Delta \rightarrow \Delta - \pi/2$ \figrefp{fig:int_schem}{(a)(iii)}. In our analytic treatment, this means that the norms of the outgoing waves
\begin{equation}
  T_\pm = \frac{1 \pm \cos(\Delta)}{2}.
\end{equation}
The predicted and computed $\Delta$-dependence of $T_\pm$ is shown in \figreft{fig:int_schem}{(b)}. The skew with respect to the analytic prediction we quantify in \secreft{section:two_soliton_collision} corresponds to the nonlinear enhancement of the phase-dependence reported in Ref.~\cite{martin_ruostekoski_arxiv_2011}.

To conclude, we have presented a general and detailed analysis of the collision of two fast-moving bright solitons at a narrow potential barrier in the NLSE. We have developed an analytic treatment of this problem, based on the assumption of a $\delta$-function potential and short collision times. Our numerical simulations of the same problem reveal that this analytic treatment is quantitatively accurate in the limit of narrow barriers and fast solitons as described in \secreft{subsec:one_soliton_Gaussian}. At realistic soliton speeds and barrier widths, however, our numerical results differ from the analytic prediction; we have quantified this in terms of the phase-skew $\epsilon$. Our analytic treatment also provides an estimate of the regimes in which the outgoing waves contain solitons. One important application of our analysis is describing the operation of a bright solitary matter-wave interferometer in a ring trap. However, we stress that our analysis remains general, and could potentially be used to describe a range of possible interferometry experiments, either in bright solitary matter-waves or other physical systems.

\section{Acknowledgments\label{section:ack}}

We thank S. L. Cornish, F. Cattani and P. M. Sutcliffe for useful discussions and the UK EPSRC (grant no. EP/G056781/1) for support. TPB also acknowledges support from Durham University.



%

\end{document}